\documentclass[twocolumn,prd,letterpaper,unsortedaddress,superscriptaddress,epsfig]{revtex4}

\usepackage{amssymb, amsmath, graphicx}

\begin{document}

\title{Ultra-Relativistic Magnetic Monopole Search with the ANITA-II Balloon-borne Radio Interferometer}

\author{
M.~Detrixhe$^4$,
D.~Besson$^4$,
P.~W.~Gorham$^1$,
P.~Allison$^1$,
B.~Baughmann$^3$,
J.~J.~Beatty$^3$,
K.~Belov$^9$,
S.~Bevan$^8$,
W.~R.~Binns$^5$,
C.~Chen$^6$,
P.~Chen$^6$,
J.~M.~Clem$^7$,
A.~Connolly$^8$,
D.~De~Marco$^7$,
P.~F.~Dowkontt$^5$,
M.~A.~DuVernois$^1$, 
C.~Frankenfeld$^4$,
%R.~C.~Field$^6$,
%D.~Goldstein$^2$,
E.~W.~Grashorn$^3$,
%C.~Hast$^6$,
D.~P.~Hogan$^{4,10}$,
%C.~L.~Hebert$^1$,
N.~Griffith$^3$,
B.~Hill$^1$,
S.~Hoover$^9$,
M.~H.~Israel$^5$,
A.~Javaid$^7$,
%J.~Kowalski,$^1$
%J.~G.~Learned$^1$,
K.~M.~Liewer$^{11}$,
%J.~T.~Link$^{1,13}$,
%E.~Lusczek$^{10}$,
S.~Matsuno$^1$,
B.~C.~Mercurio$^3$,
C.~Miki$^1$,
%P.~Mio\v{c}inovi\'c$^{1}$,
M.~Mottram$^8$,
J.~Nam$^{12}$,
%C.~J.~Naudet$^{11}$,
R.~J.~Nichol$^8$,
K.~Palladino$^3$,
%K.~Reil$^6$,
A.~Romero-Wolf$^1$,
%M.~Rosen$^1$,
L.~Ruckman$^1$,
D.~Saltzberg$^9$,
D.~Seckel$^7$,
G.~S.~Varner$^1$,
A.~G.~Vieregg$^9$,
%D. Walz$^6$,
Y. Wang$^2$.
%F.~Wu$^2$
}
\vspace{2mm}
\noindent
\affiliation{
$^1$Dept. of Physics and Astronomy, Univ. of Hawaii, Manoa, HI 96822.   
$^2$Stanford Linear Accelerator Center, Menlo Park, CA, 94025.
$^3$Dept. of Physics, Ohio State Univ., Columbus, OH 43210. 
$^4$Dept. of Physics and Astronomy, Univ. of Kansas, Lawrence, KS 66045. 
$^5$Dept. of Physics, Washington Univ. in St. Louis, MO 63130. 
$^6$Dept. of Physics, National Taiwan University, Taipei, Taiwan.
$^7$Dept. of Physics, Univ. of Delaware, Newark, DE 19716. 
$^8$Dept. of Physics, University College London, London, United Kingdom.
$^9$Dept. of Physics and Astronomy, Univ. of California, Los Angeles, CA 90095.
$^{10}$Currently at Department of Physics, University of California, Berkeley, California 94720, USA.
$^{11}$Jet Propulsion Laboratory, Pasadena, CA 91109.
$^{12}$Ewha Womans University, Seoul, South Korea.
}
\collaboration{ANITA Collaboration}
%\noaffiliation

%\author[ucb]{D.P.~Hogan}\ead{hogan.danielp@gmail.com}\address[ucb]{none}

%\begin{keyword}ANITA \sep magnetic monopole \PACS 14.80.Hv \end{keyword}

\begin{abstract}
We have conducted a search 
for extended energy deposition trails left by ultra-relativistic magnetic monopoles interacting in Antarctic ice. 
The non-observation of any satisfactory candidates in the 31 days of accumulated ANITA-II (Antarctic Impulsive Transient Apparatus)
flight data 
results in
an upper limit on the diffuse flux of relativistic monopoles. We obtain a 90\% C.L. limit of order $10^{-19}(\text{cm}^2\text{ s sr})^{-1}$ for values of Lorentz factor, $\gamma$, $10^{10}\leq\gamma$ at the anticipated energy $E_\text{tot} =10^{16}$ GeV.  This bound is stronger than all previously published experimental limits for this kinematic range.

\end{abstract}

\maketitle

\section{Magnetic Monopoles}
The search for magnetic monopoles has been the focus of 
concerted experimental effort since Maxwell's original formulation of electromagnetism 150 years ago\cite{Maxwell65}. The non-observation of an obvious partner to electric charge is somewhat problematic, although inflation provides a 
mechanism for diluting the primordial monopole abundance to miniscule densities in the current epoch\cite{Guth81}.
Searches have spanned a wide range of monopole masses and Lorentz gamma; all searches assume the Dirac angular momentum quantization condition relating the unit of electric charge $e$ to the unit of magnetic charge $g_M$, via
$2eg_M/c=\hbar$\cite{Dirac31}. %MONOPOLE WIKI PAGE
Thus far, no report of magnetic monopole detections\cite{Price75, Cabrera82, Caplin86} have been substantiated\cite{Price78, Caplin86, Huber90}. In 1970, Parker pointed out that the abundance of magnetic monopoles is constrained by the requirement that magnetic monopole currents be insufficient to deplete galactic magnetic fields\cite{Parker70}.  Only in the past decade have astrophysical experiments been able to improve upon the original Parker flux bound of $\sim 10^{-15}(\text{cm}^2\text{ s sr})^{-1}$\cite{Ambrosio02}.  The first observational astrophysics experiment to obtain limits stronger than the Parker bound was MACRO. For monopole velocities $v =\beta c>0.99c$, MACRO obtained a flux upper limit of $1.5\times10^{-16}(\text{cm}^2\text{ s sr})^{-1}$.  Upper bounds of this order of magnitude were also reported for $4\times10^{-5}<\beta<0.99$\cite{Ambrosio02}.  Subsequently, stronger bounds were reported by the PMT-based neutrino telescopes AMANDA\cite{Wissing07} and Baikal\cite{Aynutdinov05, Baikalnote}, although these latter searches target values of $\gamma$ somewhat lower than for the search described herein.

The SLIM experiment at the Chacaltaya High Altitude Laborary in the mountains of Boliva offers sensitivity to so-called
Intermediate Mass Monopoles (IMMs).  This nuclear track detector experiment is designed especially to search for monopoles (mass $10^5-10^{12}$ GeV) over a wide range of velocities (including $\beta\geq4\times10^{-5}$ for 1-Dirac-charge monopoles).  SLIM's latest monopole flux limit at $\beta\approx1$ is $6.5\times10^{-16}(\text{cm}^2\text{ s sr})^{-1}$ for Earth-crossing monopoles, or $1.3\times10^{-15}(\text{cm}^2\text{ s sr})^{-1}$ if upgoing monopoles are absorbed by the Earth\cite{Balestra08}.
Stronger flux limits based on astrophysical considerations (such as an ``extended Parker bound'') of less than $3\times10^{-22}(\text{cm}^2\text{ s sr})^{-1}$ for IMMs, using an updated model of galactic magnetic fields\cite{Adams93} have also been proposed. \message{EXPAND THIS TEXT}

\subsection*{Relativistic IMMs}
Because of their moderate mass, IMMs may acquire highly relativistic velocities.  Wick \emph{et al.}\cite{Wick03} use a model of monopole traversal of intergalactic magnetic fields (similar to the model underlying the Parker bound) to estimate that IMMs created in the early universe would now have typical kinetic energies on the order of $10^{16}$ GeV.  PeV-mass monopoles would therefore reach Lorentz factors $\gamma\approx 10^{10}$.
The fact that IMMs acquire such ``ultra-relativistic'' $\gamma$ values provides a mechanism for their detection.  Any particle traveling through a medium loses energy, but ultra-relativistic charged particles do so dramatically, initiating bright showers\cite{Jackson62}.  As noted by Wick \emph{et al.}\cite{Wick03}, experiments based on in-ice radiowave detection of showers are particularly well-suited to ultra-relativistic monopole detection because of a combination of large effective volume and favorable scaling with energy.  
Most recently, the RICE experiment 
translated their non-detection of radio emissions
from compact neutrino-induced in-ice showers into a limit on the relativistic 
monopole flux using five years of accumulated data\cite{Hogan08}. This was
not done by a dedicated search, per se, but rather by calculating the efficiency, from Monte Carlo simulations, for monopole-induced showers to pass their
neutrino search requirement criteria.
It is through detection of such bright showers that ANITA is sensitive to magnetic monopoles.

%\section{Light Relativistic Monopoles}Although GUT-scale monopoles are commonly believed to be extremely heavy ($\sim10^{17}$ GeV \cite{Vilenkin94}) and undetectably rare as a result of inflation\cite{Guth81}, there are other mechanisms resulting in production of much lighter monopoles after inflation.  Wick et al.\cite{Wick03} summarize these mechanisms, which rane from novel symmmetry breaking to a lower-than-anticipated GUT scale.  In fact, these lighter monopoles have been suggested as possible ultra-high energy cosmic rays  beyond the GZK\cite{GZK} cutoff.  Such considerations motivate the search for so-called intermediate-mass monopoles (IMMs), that is, monopoles with masses from about $10^5$ GeV to about $10^{13}$ GeV, significantly less than the conventional GUT energy.  An order-of-magnitude estimate has shown that such monopoles would acquire typical kinetic energies on the order of $10^{16}$ GeV\cite{Wick03}, giving them ultra-relativistic velocities.  Our study concentrates on this mass range because such monopoles are expected to be readily detectable by the ANITA array. Crucial to the detection scheme is an accurate model of the energy deposition as monopoles traverse matter.

\section{ANITA}
The Antarctic Impulsive Transient Antenna (ANITA) is a balloon-borne antenna array primarily designed to detect radio wave pulses caused by neutrino collisions with matter, specifically ice. The basic instrument consists of a suite of 40 Seavey Corp. 
quad-ridged horn antennas, optimized
over the frequency range of
200-1200 MHz, with separate outputs for
vertically vs. horizontally incident radio frequency signals, mounted on a high-altitude balloon.
From an elevation of
$\sim$38 km, the balloon scans the
Antarctic continent in a circumpolar trajectory.
After launching from McMurdo Station, ANITA-II was
aloft for a period of 31 days with a typical instantaneous
duty cycle exceeding 95\%.

\subsection{Detector Description}
Details on the ANITA hardware, as well as the triggering scheme crucial to the analysis described herein, are provided elsewhere\cite{ANITA_instrumentation_paper}. We here provide a brief summary of the hardware elements essential to this analysis.

The front-end Seavey quad-ridge antennas provide the first
element in the ANITA radio wave signal processing chain. These horn antennas
have separate Vertical (VPol,
defined as the zenith direction)
vs. Horizontal (HPol) signal polarizations, and each have a field-of-view
of approximately $60^o$ full-width-half-maximum (FWHM) in 
both azimuth angle ($\phi$) and zenith angle 
($\theta$). Two contiguous antennas in azimuth
define a logical `phi sector'. Following the Seavey antennas,
initial signal conditioning restricts the frequency bandpass to that
desired for the primary neutrino search, namely, 200--1200 MHz.
Signals are then split into a `trigger' path
consisting of a heirarchy of trigger conditions
(``levels'') and a `digitization' path; if all levels
of the trigger are satisfied, the digitized signals are then
stored to disk.

\subsection{Summary of ANITA missions to date}
Hadronic and electromagnetic showers resulting from
in-ice neutrino interactions produce a coherent, radio frequency Cherenkov radiation signal. 
Two one-month long missions (ANITA-1; Dec. 2006-Jan. 2007 and ANITA-II; Dec. 2008-Jan. 2009) have yielded the world's-best limits to the ultra-high energy
(UHE) neutrino flux in the energy range to which
ANITA is sensitive\cite{ANITA1-jiwoo,ANITA2-abby}. A recent analysis of the ANITA-1 data sample has
also provided a statistically large (16 events) sample of
self-triggered radio frequency signals attributed to
geosynchrotron radiation associated with cosmic-ray
induced extensive air showers
(EAS)\cite{ANITA1-stephen-EAS}. 
The analysis described herein is based on the ANITA-II data sample.
%That analysis not only demonstrated that radio wave detectors can independently trigger on EAS, but also established the air-borne strategy as an efficient EAS detection technique.

\section{Monopole Energy Loss in Matter}
\label{sec:eloss}

Essential to this monopole search is a reliable simulation of in-ice showers
caused by monopoles traversing the ice sheet.
Our model of monopole energy loss is based on the muon/tau energy loss model of Dutta \emph{et al.}\cite{Dutta01}.  In this model, energy loss by a muon traversing a medium is expressed as
\begin{equation} \label{losssummary}
-\frac{dE}{dx} = \alpha + \beta E.
\end{equation}
The $\alpha$ term is the energy loss per distance due to ionization of the medium.  The $\beta$ term\cite{notebeta} is the sum of three terms reflecting bremsstrahlung, pair production, and photonuclear effect energy losses. 
The values of $\alpha$ and of the various contributions to $\beta$ are only weakly $\gamma$ dependent.

Although energy loss due to ionization can be treated as smooth and continuous with little loss of accuracy, the stochastic fluctuations in pair production and photonuclear energy losses must be explicitly modeled; such discrete processes, in fact, provide the catastrophic showers to which ANITA-II is most sensitive.  The Dutta \emph{et al.} model expresses these energy losses in terms of partial interaction cross sections with respect to interaction energy, making it possible to isolate the expectation number of particle/medium interactions at a given energy and replace it with a random number of interactions drawn from the appropriate Poisson distribution.
To convert the stochastic model of muon energy loss to a model of magnetic monopole energy loss, the muon mass must first be replaced by the magnetic monopole mass.  For monopoles, bremsstrahlung is negligible and is disregarded\cite{Wick03}.  
Next, due to Dirac's quantization condition, a magnetic monopole of 1 Dirac charge will lose energy equivalent to an electric charge of $1/(2\alpha)$ times the proton charge\cite{Jackson62}.  Accounting for this large effective charge only
requires multiplying the expectation number of interactions by $1/(2\alpha)^2 \approx 4700$.

\begin{figure}
\centering
\includegraphics[width=0.5\textwidth]{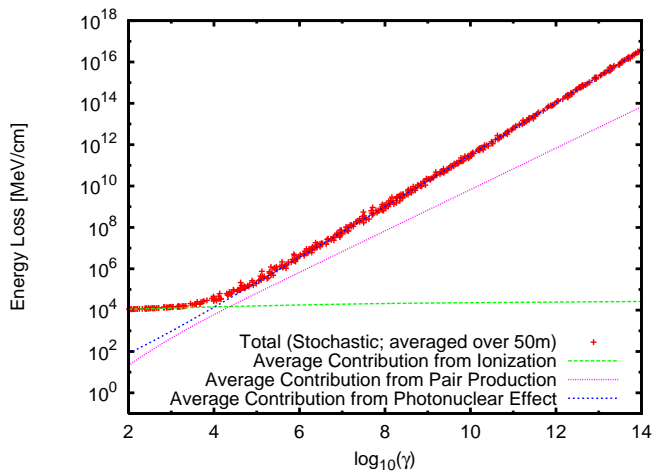}
\caption{Total energy loss versus $\gamma$ for $10^{16}$ GeV monopoles, showing stochastic variation averaged 
over 50 m intervals.  Lines show average contributions from different processes.}
\label{monobyparts}
\end{figure}

Figure \ref{monobyparts} shows various contributions to the energy loss of a  $10^{16}$ GeV monopole.  
(The monopole rest mass is constrained to vary inversely with gamma such that total energy is fixed at the reference energy of $10^{16}$ GeV.)
The curves indicate average energy loss due to the three principal mechanisms, while the points show actual stochastic energy loss (as averaged over a 50 m interval in our simulation).  
The photonuclear effect is the dominant energy loss mechanism at $\gamma>10^4$, while ionization energy losses dominate below this value.  Because the photonuclear mechanism results in hadronic showers generated by nuclear recoils, we may ignore Landau-Migdal-Pomeranchuk (LPM)\cite{LPM} effects.

\section{Monopole Simulation}
\label{sec:MC}
The monopole energy loss model described above has been incorporated into a Monte Carlo simulation of a magnetic monopole traveling within the ice volume visible to ANITA.
The Monte Carlo simulation randomly generates a monopole trajectory (uniform over $4\pi$) and energy loss according to the previously
described model, then determines the voltage response of each of the 
ANITA antennas which participate in the event trigger, as summarized below and elsewhere.
Earth curvature effects, which are considerable at the largest radii ($\sim$600 km), are incorporated into the model by a suitable adjustment of the refracted ray geometry, relative to the balloon, for a given source location. 

\subsection{Passage through Earth}

As can be inferred from Figure \ref{monobyparts}, a down-coming monopole having $\gamma\sim 10^9$ will easily traverse the entirety of the Antarctic ice sheet without substantial degradation of energy, whereas a normally-incident up-coming monopole at those $\gamma$ values will typically range out in the Earth before reaching the South Pole.
Upcoming monopoles with larger $\gamma$ values will generally reach the detector, although their energy loss in-transit through the Earth can be substantial.  For calculating this energy loss, the terrestrial density integrated over the length of the monopole's path [g/cm$^2$], as a function of approach angle, is taken from the Preliminary Reference Earth Model\cite{Dziewonski81}.  

%Figure \ref{earth_gamma} shows the energy remaining after crossing the Earth for monopoles initially having energy $10^{16}$ GeV with a range of incidence angles.  Energy loss increases with $\gamma$, so monopoles with $\gamma\lessapprox 10^7$ lose a negligible fraction of their incident energy.  At higher $\gamma$, energy loss can be significant.    Beyond $\gamma\gtrapprox 10^{10}$, the Earth is opaque to monopoles.
%\begin{figure}\centering \includegraphics[width=0.5\textwidth]{earth_gamma_standard}\caption{Final energy of magnetic monopoles with initial energy $10^{16}$ GeV after crossing the Earth.  The angle labeling each curve is the opening angle between the monopole's velocity vector and the zenith at the monopole's point of exit.}\label{earth_gamma}\end{figure}

\subsection{Propagation Through Ice} \label{Propagation}
\begin{figure}
\centering
\includegraphics[width=0.5\textwidth]{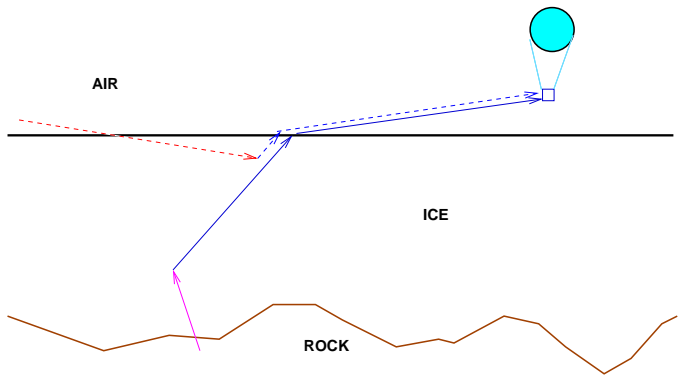}
\caption{Path geometry of monopoles coming from above (air$\to$ice) or below (rock$\to$ice) illuminating ANITA. Red/pink arrows
indicate monopole trajectories which can result in a significant signal measured at the balloon. For isotropic monopole flux, the fraction of monopoles satisfying this geometry is of order $\sim$1\%.
(For the sake of illustration, ray curvature through the firn has been neglected, although this feature is incorporated into our simulation.)}
\label{fig:mono.xfig.eps}
\end{figure}

After calculating the energy lost by a monopole en route to the ice target
(zero if the monopole arrives from above), the monopole's interaction with the ice itself is simulated,
over discrete step sizes of 15 cm.
%So long as the monopole is ``out of range'' of the ANITA antennas, it is propagated along in 10 m steps, with its $\gamma$ value decremented in each step to account for energy lost in traveling that 10 m path.  A monopole is considered to be ``in range'' as long as at least one of the ANITA antennas is within 0.36 rad of the Cherenkov cone anywhere along a 1 km m segment extending from 500 m ahead of the monopole to 500 m behind it; this requirement corresponds to $\sim 2.5\sigma_{\theta_C}$ at 200 MHz, which is the lowest sensitive frequency of the ANITA antennas. We exclude Cherenkov signal contributed beyond the 0.36 radians angle, which, although weaker in strength than the signal detected at $\theta_{Cherenkov}$, potentially offers some enhancement in aperture.  
This selection of step size is motivated by several considerations: a) the time delay between signals received at the ends of the step should be smaller than the time scale of the trigger logic, which is determined by the 7 ns duration of the front-end tunnel-diode signal integration, b) the time delay between successive steps should not be larger than the ANITA digitization time of 0.4 ns, c) the steps should not be so fine that the simulation processing time (and array sizes) become prohibitive. For the most unfavorable geometry (a shower receding from ANITA along the line-of-sight), the time delay between successive steps is the sum of the time for the monopole to move 15 cm in ice plus the additional signal transit time in-air, or approximately 1 ns. However, this geometry results in a Cherenkov pattern undetectable at the payload -- similarly, monopoles entering the ice at normal incidence from above ($air\to ice$) do not illuminate the balloon. Signals typically result when an up-coming monopole transits from rock to ice or a down-coming monopole approaching the balloon is incident on the air-ice interface from above at a near-glancing angle, as illustrated in Figure \ref{fig:mono.xfig.eps}. At the Cherenkov angle of approximately 1 radian, the latter results in zero time delay between successive steps, growing as the azimuthal angle relative to the balloon increases from zero. The former results in a maximum time delay of approximately one time bin between successive steps.
The geometric aperture of ANITA to monopoles traversing the ice can be crudely estimated as follows. For a Cherenkov cone developing in the ice, the solid angle illuminating the balloon is approximately the product of the
vertical width in $\theta$ ($\sin\theta d\theta\sim$0.2 rads at the lowest frequencies accessible to ANITA) times the
horizontal width in $\phi$ ($d\phi\sim$0.5 rads), giving a maximum geometric acceptance of $0.1/4\pi$, or approximately 1\%. 
This provides the greatest limitation on ANITA's sensitivity to monopole signals.

% As a result of this small step size, all the energy lost within the interval can be treated as originating at a single ``subshower'' with a point-like source, while introducing signal arrival timing errors no greater than 0.4 ns.  By comparison, the actual experiment's digitizer samples every nanosecond. 

\subsection{Simulated Antenna Response}
A Monte Carlo simulation of the radio frequency signals caused by cascades (discussed in Ref.\cite{Rice3}) previously developed in the context of RICE's high-energy neutrino flux studies has been adapted for application to ANITA. In addition to the stochastic energy loss processes enumerated previously, we also simulate ray-tracing through the firn, as well as signal loss in transmission across the snow surface into the air, as a function of incident angle on the ice-air interface.  %Thermal noise is simulated by sampling from a Gaussian distribution of width $V_{rms}=\sqrt{4GZk_BB(T_{env}+T_{sys}+T_{galaxy})}$, for which $G$ the total system gain, $k_B$ is Boltzmann's constant, $B$ is the bandwidth of each individual ANITA trigger band, $Z$=50 $\Omega$, and $T_{ice}$, $T_{sys}$, and $T_{galaxy}$ the contributions to thermal noise i) looking into the ice ($T_{ice}\sim -$223 K), ii) from the receiver system itself ($\sim$200 K), and iii) from the galaxy itself. The latter contribution is fairly negligible over the ANITA bandpass. The thermal noise voltage is generated with a random phase relative to the monopole signal and the two voltages then added in the time domain. This simulation has been calibrated against two other ANITA simulation codes used for the primary neutrino analysis, and found to be in good agreement.

\begin{figure}
\centering
\includegraphics[width=0.5\textwidth]{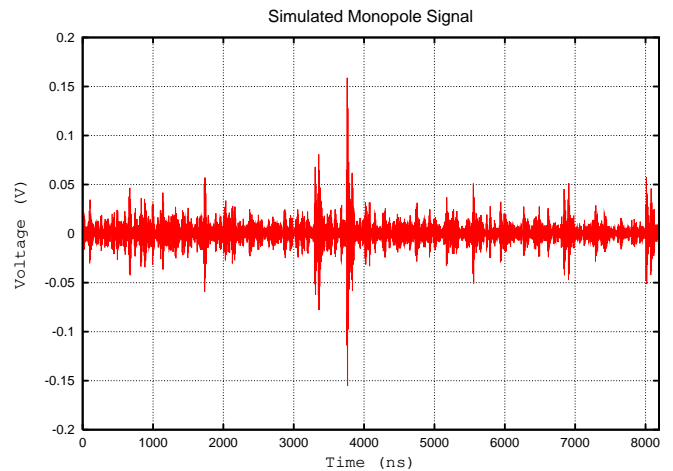}
\caption{Simulated voltage vs time (as measured at the data acquisition system) in a single ANITA vertically-polarized (VPol) antenna channel;
$\gamma=10^9$.}
\label{voltgraph}
\end{figure}

A typical example of a simulated monopole time-domain
voltage profile, for a monopole simulated at $\gamma=10^9$, is shown in Figure \ref{voltgraph}. In contrast to temporally compact showers from neutrinos, monopoles deposit energy over a timescale considerably longer than the 
minimum $\sim$400 ns time required to fill four available data acquisition event buffers. Schematically, the monopole signal can be treated as a sequence of `subshowers', each having a characteristic energy deposition geometry and time delay relative to ANITA. As demonstrated by laboratory 
measurements\cite{slac-testbeam,slac-salt04}, the frequency-domain radio wave
signal due to a shower in a dense medium is typically broad and, in contrast
to typical narrow-band anthropogenic backgrounds, extends over hundreds of MHz.
The complete simulated voltage profile $V(t)$ at each antenna is determined by coherently summing the voltage contributions of various subshowers in each time bin.  The signal phase at the emission point will vary slowly with viewing angle\cite{buniy}; however, in most cases the dominant emission arises from a coherent region along the track centered around the Cherenkov point and subtending a small viewing angle.  Accordingly, we ignore all signal phases other than those from travel time in our analysis.

Similar to showers from neutrinos, the signal strength is typically at least an order of magnitude stronger along the vertical polarization
(VPol) axis compared to the horizontal polarization axis, given the excellent cross-polarization isolation ($\sim$20 dB) of the Seavey antennas.

%http://www.phys.hawaii.edu:8080/anita_notes/411
\subsection{Trigger Simulation}
The heirarchical ANITA-II trigger allows a
relatively low neutrino detection trigger threshold, while maintaining
a tolerable ($\sim$Hz) thermal noise data rate written to disk.
In contrast to ANITA-I, only
signals from the VPol channel of the
dual-polarization horn antennas contribute to the 
ANITA-II trigger.
Following the antenna, signals routed through the
`trigger' (vs. `digitization') path are tested for
their spectral power in four frequency bands, 
approximately covering
the intervals (in MHz) 200$\to$350, 330$\to$600, 630$\to$1100
and 150$\to$1240, respectively.
%ANITA-1: with centers at $\nu_c = $265, 435, 650, and 990~MHz and bandwidths of $\Delta\nu =$130, 160, 270, 415~MHz, or fractional bandwidths $\Delta\nu/\nu_{c} \simeq 44\%$ on average.
This partitioning is performed in order to
provide rejection power against narrow-bandwidth
anthropogenic backgrounds, while retaining high
efficiency for sharp duration (temporally), large-bandwidth neutrino signals. The frequency-banded signals are then passed through a tunnel-diode, which
integrates roughly 7-ns units of data and provides a unipolar (negative) output
pulse. The lowest-level trigger (L0) requires signal in one of the four
frequency bands at a level exceeding approximately 2.3$\sigma_V$,
with $\sigma_V$ the rms of the typical tunnel-diode 
output voltage at this point. %Coincidences of single-band triggers formed the second (L1) antenna-level stage, however, 
The L1 trigger required two of the three frequency bands plus a full-band trigger within a 10 ns window. (In an improvement over ANITA-1, the different group delays of the bandpass filters from the single-band stage were compensated digitally to ensure synchronization.)
The L1 triggers were then combined for the L2
trigger stage. For the upper and lower antennas, an L2 was issued in one of the 16 phi sectors when two of the three antennas either on or adjacent to that phi sector triggered within approximately 10 ns of each other. For phi sectors containing a nadir antenna, an L2 was issued for each L1, and for those without, an L2 was issued when either adjacent nadir antenna received an L1. Finally, the fourth (L3) stage required two of the three antenna rings (upper, lower, and nadir) to have an L2 trigger within approximately 10 ns. Satisfaction of all trigger tiers initiates digitization of all antenna channels and occurs at a rate of approximately 10 Hz. By contrast, the L1 and L2 rates are approximately 2--3 MHz, and 2.5 kHz per antenna pair, respectively. Laboratory studies have demonstrated that trigger efficiency differences between ANITA-1 and ANITA-II result primarily from the removal of the HPol antenna channels from the trigger logic, and the modified frequency banding.

\section{ANITA Experimental Monopole Detection Strategy}

\label{Event Signature}
\begin{figure}
\centering
\includegraphics[width=0.5\textwidth]{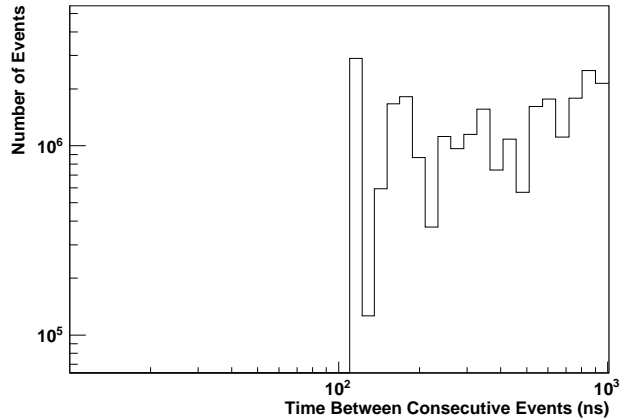}
\caption{Minimum time between successive triggers (in-flight data).}
\label{minTimeFigure.eps}
\end{figure}

As a monopole interacts with the ice, it sheds energy stochastically, so the monopole signal will be comprised of a string of impulsive subshower events. When ANITA triggers, the signal is temporarily stored in one of four buffers and is then read to hard disk. 
As soon as one buffer is filled, it is
read out and then reset for the next event. Since this process requires
some minimum time, if a series of 
triggers occur in rapid succession (e.g., within 1 $\mu$s), there is
insufficient time to maintain at least one free buffer. Once all
four buffers fill, read, clear, and reset commands are sequentially issued,
requiring a minimum time
of tens of milliseconds, 
much longer than the monopole signal would be visible to the balloon.
The experimental monopole
signature registered in ANITA is therefore expected to consist of the first four
threshold-crossings ($\sim$500 nanoseconds total) once the monopole
comes into view; the remaining signal produced by the monopole ionization
trail is not registered during the dead-time
required after the four
buffers initially fill.
The 500 nanosecond estimate also includes
a small delay between successive triggers, as they fill the available buffers. The experimental in-flight data show a minimum time between triggers $T_{min}$=112 ns (Fig. \ref{minTimeFigure.eps}), to be compared with the 100 ns of data recorded
in a standard waveform capture. This value is commensurate with laboratory measurements prior to the flight. 

\section{Monopole Selection and Event Yield}
\begin{figure}
\centering
\includegraphics[width=0.5\textwidth]{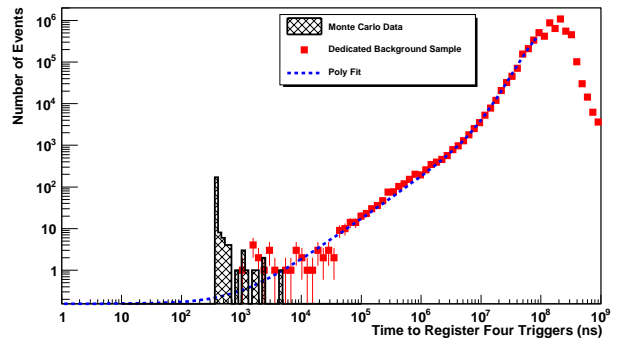}
\caption{$T_4$ distribution for Monte Carlo monopole
signal simulations compared with DB data (points). 
Note that, for our simulations, we 
conservatively use $T_{min}$=120 ns.
Also overlaid is a fit to the DB data showing the extrapolation into the region $T_4<$500 ns, used to estimate the expected anthropogenic background for the signal sample.}
\label{fig:MCAnitaPretty}
\end{figure}

Our monopole event selection criterion therefore consists of four `fast' triggers over a time
interval $T_4$; it remains only to define how
quickly the triggers must be registered ($T_{max}$) in order to be considered a monopole candidate. 
To minimize bias, $T_{max}$ was determined using
a `dedicated background' (DB) subset of the total ANITA dataset, defined as all data taken when the balloon was within 300km of McMurdo base. This subset comprised approximately 15\% of the total data. For this dataset, Figure \ref{fig:MCAnitaPretty} shows the $T_4$ distribution, with the monopole MC simulated distribution overlaid, as well as a polynomial fit to the DB distribution.
For large values of $\gamma$, the stochastic nature of the monopole energy loss tends to induce threshold crossings at a rate much faster than that at which the instrument can trigger. Since the filling of all buffers will immediately incur tens of milliseconds of subsequent dead time, the simulated $T_4$ distributions therefore exhibit pile-up at the minimum possible value of 3$T_{min}$.

From the DB sample, we set the requirement that $T_4<T_{max}~(\equiv$ 500 ns), which eliminates all the background event triggers. This is the main monopole selection criteria then applied to our `search' sample.

\subsection{Trigger Efficiency}
Monte Carlo simulations of monopoles isotropically illuminating the 
Antarctic ice sheet are used to determine the $T_4<$500 ns efficiency.
Although the bulk of high-energy monopoles which produce four triggers also
satisfy the $T_4<$500 ns criterion, the limited
geometric acceptance for producing four triggers ($<$1\%, designated as
``$\epsilon_a$'') provides the main
limitation on the total efficiency (designated as
``$\epsilon_t$'').
The ratio of events satisfying both the simulated
trigger (which includes aperture), 
as well as the $T_4<$500 ns requirement 
relative to the
total events generated in simulation for each $\gamma$ value
(typically 20,000), are as follows:
\begin{center}
\begin{tabular}{|c|cccccc|} \hline
$log_{10}(\gamma)$ & 8 & 9 & 10 & 11 &12 & 13 \\ \hline
%$N_{events}^{simulated}$ & 2157  & 1562 &   7785 &   10519+12236 &  3799  &   4370 \\
%$N_{events}^{triggered}$ & 0   & 2   & 23  &  42  & 19  & 26 \\
$\epsilon_a$ ($\times 10^{-3}$) & 0 & 0.43 & 3.85 & 4.35 & 7.19 & 7.69 \\
$\epsilon_t$ ($\times 10^{-3}$) & 0 & .06 & 1.65  & 3.17  &   4.66  &   6.12 \\ 
$\sigma(\epsilon_t)$  ($\times 10^{-3}$) & 0 & .04 & .34 & .48 & .99 & 1.20 \\
\hline
\end{tabular}
\end{center}
Also shown in the last row are the errors on the total efficiency; our
eventual limits will later be degraded by one statistical error bar in each energy
bin.

\subsection{Background estimate in search sample}
After establishing the requirement that $T_4<$500 ns
based on the DB sample, the trigger time distribution of the
remainder of the data set (the monopole `search' sample) was considered. 
It is desirable 
that the number of events passing this requirement in
the search sample be small
enough that they can be individually 
hand-scanned and their candidacy assessed against
Monte Carlo simulated monopole templates.
Extrapolating the DB $T_4$ distribution into the `signal' region
below 500 ns (Fig. \ref{fig:MCAnitaPretty}) 
indicates a projected yield of 0.46 events; correcting
this by the ratio of livetime in the search sample relative to the
DB samples (0.85/.15) results in an {\it a priori} expectation
of 2.6 background events in the search sample.

Thermal
triggers, overall, dominate the accumulated ANITA-II data sample and are responsible for the `hump' to the right
in Figure \ref{fig:MCAnitaPretty}.
Given the typical L3 trigger rate of 10 Hz, the probability of registering 4 uncorrelated thermal triggers within 500 ns can be estimated to be $\sim (5\times 10^{-6})^3$ per second, implying a negligible
expected thermal
event fake yield in our search sample.

\subsection{$T_4$ distributions in search sample}
\begin{figure}
\centering
\includegraphics[width=0.5\textwidth]{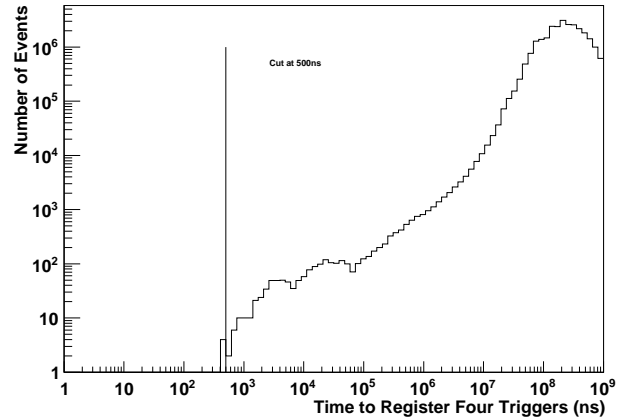}
\caption{Time between first and fourth event trigger ($T_4$) distribution for ANITA-II monopole search sample; the monopole
event selection requirement ($T_4<$500 ns)
is indicated by vertical line.}
\label{fig:fullDataPretty}
\end{figure}

The $T_4$ distribution for the entire ANITA-II flight is shown in Figure \ref{fig:fullDataPretty}. 
We do note dissimilarities, at small values of $T_4$, in the shapes of the DB vs. search sample $T_4$ distributions. 
This is, however, not necessarily surprising, since our DB sample is restricted to data taken in the vicinity of
McMurdo Station, while the search sample includes data taken over the entire continent.

\begin{figure}
\centering
\includegraphics[width=0.5\textwidth]{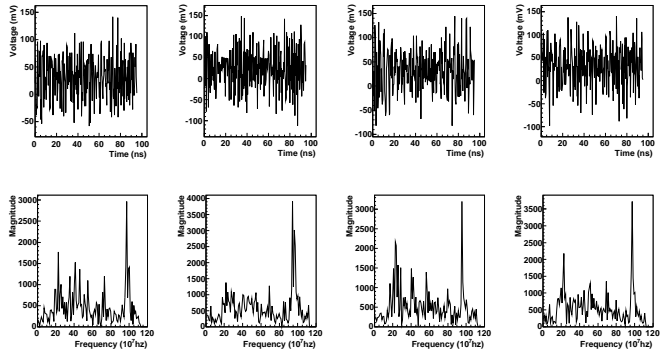}
\caption{Time domain (top four panels), as well as frequency domain
(bottom four panels) voltage distributions for the four channels in
one of the four search sample events passing the $T_4<$500 ns requirement.}
\label{fig:BK}
\end{figure}

\begin{figure}
\centering
\includegraphics[width=0.5\textwidth]{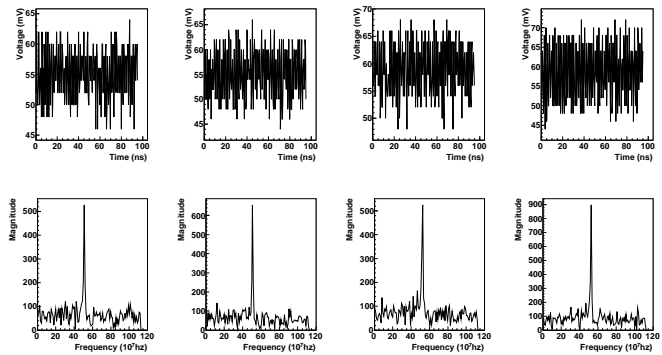}
\caption{Time domain (top four panels), as well as frequency domain
(bottom four panels) voltage distributions for the four channels in
one of the DB events
with $T_4\approx$500 ns.}
\label{fig:DBBK}
\end{figure}

In the entire data sample, only four events contain four
rapid triggers which satisfy the 500 ns maximum total trigger time criterion. All four of these events were already classified as background, using the rejection criteria developed for the primary ANITA-II neutrino search\cite{ANITA2-abby}. Specifically, the first two events were rejected on the basis of their temporal near-coincidence (occurring within 15 seconds of each other) as well as having source locations (after filtering for strong continuous wave [CW] components)
consistent, to within $\sim$1 degree, with the South Pole.
The third event was recognized as payload noise, on the basis of an anomalous DC offset, and saturated waveforms in all phi sectors.
The final event was rejected on the basis of its power spectrum, which is observed to be dominated by one narrow CW frequency without which the event would be sub-threshold for analysis. Removing that one line from the frequency spectrum results in a time-domain waveform which also reconstructs to a known background source location in West Antarctica and further affirms this as a background event. 
The time-domain, as well as frequency-domain distributions for the four highest-amplitude channels in that last event
are displayed in Figure \ref{fig:BK}.
We note similarity in character between that event from the search sample
and a typical event with $T_4\approx$500 ns from the DB sample (Figure
\ref{fig:DBBK}).
In fact, the power spectra of all the events with $T_4<$500 ns, 
in both the DB as well as search samples, are found
({\it a posteriori}) to be dominated by CW lines. This is in marked contrast to
monopole signals, which exhibit the broadband characteristics of temporally-sharp,
coherent radio wavelength signals. 

Elimination of these four events in the search sample (roughly consistent with our {\it a priori} expectation of 2.6 background events from the DB sample) results in zero candidate monopole triggers in the ANITA-II data set.
Since no satisfactory event candidates are found, we therefore choose
to set a limit on the monopole flux.
However, having invoked criteria (base rejection) used for the primary neutrino search, we also must
later degrade our monopole detection efficiency by the geometric loss due to that anthropogenic
background requirement ($\epsilon_b=0.63$).

\section{Cross-check analysis}
\begin{figure}
\centering
\includegraphics[width=0.5\textwidth]{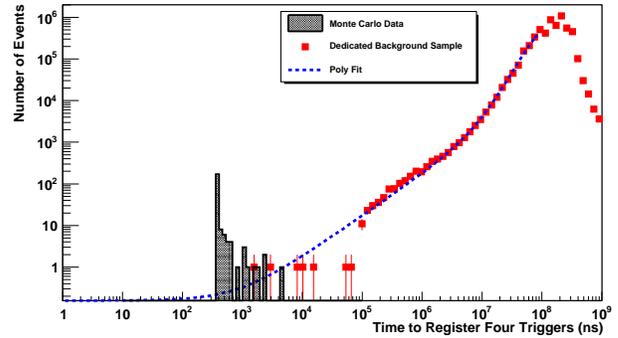}
\caption{$T_4$ distribution for Monte Carlo simulations and DB sample, for events passing $PFP<10^{-7}$ requirement. Note the
diminution of dedicated background events populating the $T_4\sim$500 ns region, in comparison to Figure 5.}
\label{MC1}
\end{figure}

A complete, and parallel analysis, which was intended to avoid the {\it a posteriori} hand scan, was performed as follows.
Given the fact that anthropogenic noise tends to be a) CW (continuous wave) in the frequency domain, and 
b) persisting for durations of order seconds, 
we initially restrict our candidate search sample to those events satisfying an extended CW-rejection requirement.
Starting with the first event of a set of four filled in rapid succession, we transform to the frequency domain and determine the frequency bin with the highest fractional power. For a uniform population, and based on 256 time-domain 
samples, this quantity would be 1/(256/2), or 1/128. Assuming that, for man-made backgrounds, that bin will also have the highest spectral power for the next 3 events in a candidate quartet, we now calculate the product of the fractional power (``PFP''), for that same bin determined from the first captured event, for all four events in a quartet. For thermal noise, this quantity peaks at around $10^{-9}$, for Monte Carlo simulations, this distribution peaks at around $10^{-8}$, primarily because of the fact that: i) monopole candidates are generally close to the thermal floor, and ii) the ANITA trigger requires power in multiple trigger bands. Based on the Monte Carlo distribution, we set a cut on this statistic at $PFP<10^{-7}$, which corresponds to approximately 95\% efficiency and high
background rejection of the DB sample (Figure \ref{MC1}).
Having set the cut based on Monte Carlo simulations 
only, we now apply this requirement to the signal search sample; none of the $T_4<$500 ns data events survives this requirement, resulting (again) in zero monopole candidates. Note that application of this CW-rejection would actually result 
in a stronger limit than what we finally quote below as we do not incur a penalty of $\epsilon_b$ in our efficiency.

\section{Live Time}
\begin{figure}\centering\includegraphics[width=0.5\textwidth]{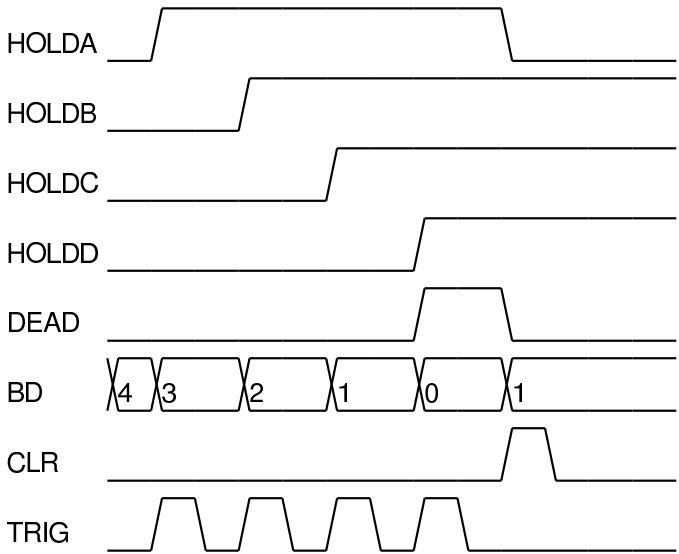} %{timingDiagram}
\caption{Timing diagram, 
illustrating the sequence of filling and emptying event data buffers.
Four rapid triggers (``TRIG'') initiate data HOLD in
buffers A, B, C and D, and decrements the Buffer
Depth (BD) variable. Once all buffers are filled, DEAD time is
incurred during the clear (CLR) and reset cycles.
\label{timing_diagram}}\end{figure}

In order to calculate a flux limit, the time ANITA is sensitive to monopoles must be known. Recall that ANITA must be both active and have four free buffers available to record our ``cluster'' of events. The buffer depth (``BD'', or the number of available buffers to fill with data triggers) was not always known during flight, so a simulation was developed to estimate the fractional live time contributed by each buffer state. We know that when an event is triggered, a buffer becomes occupied, and the buffer depth decrements (we use the notation ``BD=4'' to indicate that all buffers are empty; ``BD=0'' implies all buffers are occupied and dead time is incurred). When an event is read to disk, a buffer is freed and BD increments. Figure \ref{timing_diagram} is a graphical representation of event acquisition and readout.  

Our simulation provides an estimate of the amount of time spent in each buffer state $T_{BD=i}$, where i indicates the buffer depth; the monopole livetime therefore corresponds to $T_{BD=4}$. The only variable in the model is the unknown effective readout time $T_{readout}$, which is determined as follows. The simulation reads through the entire list of trigger times recorded in data. When an event is triggered, our model increments BD and, after $T_{readout}$ has elapsed, BD decrements. If, in the simulation, a trigger occurs when BD=0, the event is simply skipped and BD is left unchanged. Given the typical ANITA-II data trigger rate of 10 Hz, any gaps in triggers greater than 1 second are considered dead time, and following one of these gaps, BD is reset to 0. Despite the individual buffer depth states not being known, the total time ANITA was live ($T_{total} = T_{BD=1}+T_{BD=2}+T_{BD=3}+T_{BD=4}$) is reliably tracked during flight and provides a constraint on our model. The `correct' $T_{readout}$ is defined as that value for which the simulated $T_{total}$ converged on the actual $T_{total}$. 

\begin{figure}
\centering
\includegraphics[width=0.5\textwidth]{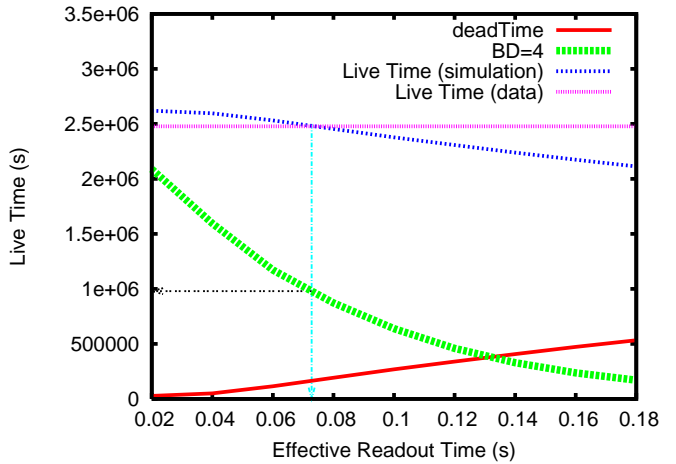}
\caption{Comparison of the simulated readout time model with data. ``Simulated'' quantities refer to outputs of the live time model, as the effective readout time is varied. At an effective readout time of 73~ms, the simulated livetime (blue, short dashed) matches the known livetime (cyan, dotted), implying a BD=4 livetime slightly less than $10^6$ s.}
\label{fig:model.eps}
\end{figure}

Figure \ref{fig:model.eps} shows the simulation results as the trial value of $T_{readout}$ is varied from 0.02s to 0.14s. 
The horizontal magenta line represents the known $T_{total}$; the intersection of this line with the cyan arrow (simulated $T_{total}$) corresponds to the case where the live time
constraint is satisfied. 
Simulation matches data when $T_{readout}\approx$73 ms. From this value,
our aggregate livetime for BD=4 is estimated to be $9.75\times 10^5$ s. 
%Not surprisingly, this value is approximately 1/4 of the total ANITA-II live time.
% -- note that this neglects any event selection effects such as our requirement that the balloon be over `quiet' ice.

\section{ Systematic Errors}
\label{sec:systematics}
The primary source of uncertainty is likely 
due to the systematic error inherent
in our efficiency estimate, primarily due to ice properties uncertainties
and our finite time binning procedure (which determines the coherence of the time
domain waveform) in simulations.
This is assessed by running our
simulation with two considerably different sets of parameters.
In the first set of trials, the monopole is tracked
over a step size of 80 cm, beginning at the first point within the ice sheet
when the monopole registers a trigger, and extending
for 8192 simulated samples. For high-energy, upcoming 
monopoles which are capable of
arriving from below after traversing the Earth, the monopole is therefore
tracked over the warmest, and most birefringent $\sim$1.5 km of ice starting at the bottom of the ice sheet.
In the second set of trials (our default parameters), the monopole is tracked over a 15 cm step size only over the top 
(colder ice, with no birefringence) 1.5 km of ice thickness, extending over 16384 simulated samples. The `test'
parameters result in approximately $\sim$23\% lower efficiency than our default parameters, largely due to the generally-smaller measured
signal strengths emanating from the deeper, warmer (and 
significantly more radio-absorbing)
ice. This extreme variation should bracket the expected uncertainties due to our radio frequencies ice properties'
parameterization, as well as our finite time-binning procedure for signal, in the limit where the step size is correctly taken to zero. 

There is some uncertainty associated with the calculation of the monopole energy loss. To assess this, we have compared our simulation with the monopole energy loss calculated independently by the {\tt mmc} code\cite{dima}, which was originally designed for muons and subsequently adapted to model monopoles for the IceCube collaboration. Unlike our simulation, the {\tt mmc} energy loss
parametrization is based on several different theoretical 
calculations, and also calculates energy losses independently of our simulations.
Since the kinematic regime targeted by the IceCube analysis is approximately 5-10 orders of magnitude lower than for this analysis, the overlap between the simulations is limited to gamma values of order $10^5$. For $\gamma=10^5$, we find agreement between the results of the 
two simulations to within 13.5\%.
%Over the range for which {\tt mmc} is designed ($\gamma\le 10^5$), we find agreement between radiomc and {\tt mmc} in both $<dE/dx>$/km, as well as shower energy distributions for 1 meter step sizes, at the 10\% level.

Our overall result is relatively insensitive to the parametrization of thermal noise in our
simulation. The addition of such noise is included in our simulation and
will, on average, lead to a noticeable efficiency improvement for
very large simulation samples. For our limited simulation samples, however, 
this effect is
mitigated by the stochastic nature of the monopole energy deposition, which
leads to a large number of in-ice showers with widely varying energies, and distributed over viewing angles
varying by up to several degrees relative to the balloon.

To account for our limited Monte Carlo statistics, we have included an error equal to the fractional statistical precision on our Monte Carlo-derived
detection efficiency. To account for the uncertainty in our choice
of $T_4<$500 ns as the optimal monopole event selection
requirement, we have also folded in a systematic error equal to the 
inefficiency of our $T_4<$500 ns event selection requirement (10\%). Added in quadrature with both the energy loss uncertainty (13.5\%), as well as the 23\% systematic error determined above yields a net systematic error of 28\%. Our final upper limit is accordingly degraded by this fraction.
Uncertainties in our livetime calculation are believed to be
smaller than 5\% and can therefore be neglected in our
total systematic error.
We note that the two dominant systematic errors above are independent of each other -- the energy loss uncertainty is assessed without fully propagating the signal to the balloon, while the latter uncertainty measures the error in our modeling subsequent to energy deposition.
%We have additionally checked our in-ice energy loss parametrization against a second monopole energy loss parametrization code (the `mmc' parametrization\cite{mmc}). 
Note that the final result is somewhat robust to changes in the overall scale of $dE/dx$ -- increasing 
$dE/dx$ leads
to brighter in-ice showers, but reduces the flux of monopoles that can penetrate to the ice sheet from below the horizon.
Conversely, reducing $dE/dx$ increases the flux arriving at the ice sheet, but decreases the magnitude of signal
strength on a monopole-by-monopole basis.
In general, ANITA's sensitivity to relativistic monopoles is 
primarily determined by the detector/ice geometry. For monopole
trajectories that do illuminate the balloon, we expect there to be
tens, if not hundreds of potential showers which can cause event triggers.
In practice, only the first four of these are recorded by the data
acquisition system. 

Overall, we have tried to take a conservative approach in calculating our
sensitivity, including: a) tracking the monopole over the only the upper
half of the ice sheet and neglecting any signal from the lower half which
results in a conservative estimate of effective area,
b) calculating sensitivity under the assumption that systematic
errors are uncorrelated, and therefore forfeiting any numerical advantage
that would be gained by taking such correlations into account,
c) using an estimated trigger efficiency which is approximately
25\% less efficient than that employed for the primary
neutrino search.

% Fig:X shows the simulated accumulated live time and the actual accumulated live time throughout the flight. There are three data sets plotted for the simulated results to show the reader the sensitivity of the simulation to perturbations in the effective readout time. Fig:X illustrates that after approximately the first hundred hours aloft, the simulated live time is within a few percent of the actual for the duration of the flight. The assumption is made that, since the live time is reasonably accurate throughout the flight, and the end result, $T_{total}$, is consistent with the experimental results, the other components ($T_{BD=1}, T_{BD=2}, T_{BD=3}, T_{BD=4}$) are approximated with reasonable accuracy. The component we are interested in specifically is $T_{BD=4}$, because it is the time ANITA is sensitive to observing monopoles by this approach. In practice, $T_{BD=4}$ is readjusted by half to account for any reasonable doubt in the accuracy of the method.  

\section{Particle Flux Limit Calculation}
Particle flux is reported in units of ${\text{cm}^{-2}\text{s}^{-1}\text{sr}^{-1}}$. To calculate an upper bound on the diffuse monopole flux, the upper limit on the number of observed particles ($N$), live time ($T_{live}$), detection area ($A$), and the solid angle from which incident particles are approaching ($\Omega$)  must be known. In this experiment, none of the candidate events met the acceptance criteria, so the number of observed particles is zero, corresponding to a Poisson 90\% confidence level upper limit on the number of observed particles of 2.3 (we neglect possible background contributions which, if included, would tend to strengthen our derived upper limit). The BD=4 live time given by the simulation is $T_{live}$=975,000 s. The detection area is the surface area of a spherical cap of radius R=680 km and is therefore slightly greater than the planar projection area of $\pi\times R^{2}$ (by approximately 5\%).
% by equation~(\ref{area}) where $D~=~680{km}~=~6.8\times10^{7}{cm}$.\begin{equation}A = \frac{\pi\times D^{2}}{4}\label{area}\end{equation}
Finally, we must include the 
efficiency for  
events to both trigger, 
as well as to pass our $T_4<$500 ns 
($\epsilon_t$) requirement, and also
the efficiency for events to pass our base rejection requirements
($\epsilon_b$) in the expression for particle flux $F$ (Eqn.~\ref{pf});
we take $\epsilon_b$ to be the value used for the neutrino-search analysis (63\%), as those search criteria are invoked during the final phase of the analysis in our elimination of the four
events passing the $T_4<$500 ns requirement from monopole candidacy.
\begin{equation}
F = \frac{N}{A\Omega T_{live}(\epsilon_t \epsilon_b)}
\label{pf}
\end{equation}

\begin{figure}
\centering
\includegraphics[width=0.5\textwidth]{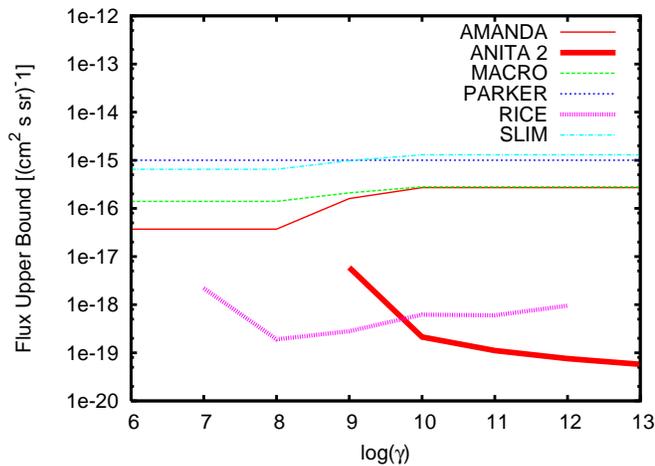}
\caption{Comparison of ANITA upper limit on diffuse monopole flux with other results. Save for RICE, 
other experimental results have been extrapolated up to our sensitive kinematic interval. In performing this extrapolation,
the limits for $\gamma\ge 10^9$ have been weakened by a factor of two, to account for increasing Earth
opacity.}
\label{fig:results}
\end{figure}

Our final results are shown in Figure \ref{fig:results}. We obtain
a flux limit of order $10^{-19}/({\rm cm^2-s-sr})$ at $\gamma=10^{10}$, improving slowly over the next few decades in monopole $\gamma$.
For large gamma values, the results for ANITA are considerably 
lower than that of any experiment to date.
Although the 
bandpasses of the RICE and
ANITA experiments are comparable, the ANITA analysis strategy is directed at monopole detection $ab~initio$, affording a factor of three improvement in efficiency relative to the RICE analysis, which was driven by a neutrino search strategy. The additional enhancement in effective collection area ($\sim$500, including losses in collection area due to rejecting bases) 
more than compensates ANITA's factor of $\sim$50 smaller livetime relative to RICE.

\section{Summary}
From the non-observation of highly ionizing showers we have derived the monopole flux upper limits shown in Fig. \ref{fig:results}, which are on the order of $10^{-19} (\text{cm}^2\text{ s sr})^{-1}$.  Previously, AMANDA\cite{Wissing07}, Baikal\cite{Aynutdinov05}, and MACRO\cite{Ambrosio02} determined monopole flux limits on the order of $10^{-16}(\text{cm}^2\text{ s sr})^{-1}$ for $\beta$ greater than $0.8$, $0.8$, and $4\times10^{-5}$, respectively.  Although the results of this study cover a much narrower range of $\beta$ values than previous works, it is the range that is of the greatest interest for IMM searches.  Within much of this kinematic range ($E=10^{16}$ GeV; $\gamma\geq10^{9}$), monopole flux limits from ANITA are stronger than the limits from any previous astrophysical monopole search.

\medskip 

{\bf Acknowledgments}
We thank the National Aeronautics and Space Administration,
the National Science Foundation Office of Polar Programs,
the Department of Energy Office of Science HEP Division,
the UK Science and Technology Facilities Council, the
National Science Council in Taiwan ROC, and especially the
staff of the Columbia Scientific Balloon Facility.


\begin{thebibliography}{99}
\bibitem{Baikalnote}The flux limit from Baikal assumes an initial monopole energy of $10^{12}-10^{15}$ GeV, lower than the initial energy assumed in this study.  Therefore, the Baikal limit shown in Fig. \ref{fig:results} should be regarded as a function of the indicated mass but not necessarily the indicated $\gamma$.
\bibitem{notebeta}The loss symbol $\beta$ should cause no confusion with the velocity $\beta$.
\end{thebibliography}
\end{document}